\documentclass[conference]{IEEEtran}
\IEEEoverridecommandlockouts
\usepackage{cite}
\usepackage{amsmath,amssymb,amsfonts}
\usepackage{algorithmic}
\usepackage{graphicx}
\usepackage{textcomp}
\usepackage{xcolor}
\usepackage{times}
\usepackage{helvet}
\usepackage{courier}
\usepackage{graphicx}
\usepackage{multirow}
\usepackage{booktabs}
\usepackage{tabularx}
\usepackage{subfigure}
\usepackage{amsmath}
\def\BibTeX{{\rm B\kern-.05em{\sc i\kern-.025em b}\kern-.08em
    T\kern-.1667em\lower.7ex\hbox{E}\kern-.125emX}}

\usepackage{fancyhdr}
\begin{document}

\title{AudioInsight: Detecting Social Contexts Relevant to Social Anxiety from Speech\\
\thanks{* The two authors contributed equally to this work.}
\thanks{\textsuperscript{\dag} Corresponding Author: Zhiyuan Wang (vmf9pr@virginia.edu)}
\thanks{This work was supported in part by a 3Cavaliers Seed Grant, by the National Institute of Mental Health of the National Institutes of Health under award number R01MH132138, and the Commonwealth Cyber Initiative, an investment in the advancement of cyber R\&D, innovation, and workforce development. For more information about CCI, visit www.cyberinitiative.org.}
}

\author{\IEEEauthorblockN{Varun Reddy*}
\IEEEauthorblockA{\textit{Department of Computer Science} \\
\textit{University of Virginia}\\
Charlottesville, USA \\
dpc3qt@virginia.edu
}
\and
\IEEEauthorblockN{Zhiyuan Wang*\textsuperscript{\dag}}
\IEEEauthorblockA{\textit{Department of Systems and Info. Engineering} \\
\textit{University of Virginia}\\
Charlottesville, USA \\
vmf9pr@virginia.edu 
}
\and
\IEEEauthorblockN{Emma R. Toner, Maria A. Larrazabal}
\IEEEauthorblockA{\textit{Department of Psychology} \\
\textit{University of Virginia}\\
Charlottesville, USA \\
\{ert6g,ml4qf\}@virginia.edu
}
\and
\IEEEauthorblockN{Mehdi Boukhechba}
\IEEEauthorblockA{\textit{Johnson \& Johnson Innovative Medicine} \\
Titusville, USA \\
mboukhec@its.jnj.com}
\and
\IEEEauthorblockN{Bethany A. Teachman}
\IEEEauthorblockA{\textit{Department of Psychology} \\
\textit{University of Virginia}\\
Charlottesville, USA \\
bat5x@virginia.edu}
\and
\IEEEauthorblockN{Laura E. Barnes}
\IEEEauthorblockA{\textit{Department of Systems and Info. Engineering} \\
\textit{University of Virginia}\\
Charlottesville, USA \\
lb3dp@virginia.edu}
}

\maketitle

\begin{abstract}
During social interactions, understanding the intricacies of the context can be vital, particularly for socially anxious individuals. While previous research has found that the presence of a social interaction can be detected from ambient audio, the nuances within social contexts, which influence how anxiety provoking interactions are, remain largely unexplored. As an alternative to traditional, burdensome methods like self-report, this study presents a novel approach that harnesses ambient audio segments to detect social threat contexts. We focus on two key dimensions: number of interaction partners (dyadic vs. group) and degree of evaluative threat (explicitly evaluative vs. not explicitly evaluative). Building on data from a Zoom-based social interaction study (N=52 college students, of whom the majority N=45 are socially anxious), we employ deep learning methods to achieve strong detection performance. Under sample-wide 5-fold Cross Validation (CV), our model distinguished dyadic from group interactions with 90\% accuracy and detected evaluative threat at 83\%. Using a leave-one-group-out CV, accuracies were 82\% and 77\%, respectively. While our data are based on virtual interactions due to pandemic constraints, our method has the potential to extend to diverse real-world settings. This research underscores the potential of passive sensing and AI to differentiate intricate social contexts, and may ultimately advance the ability of context-aware digital interventions to offer personalized mental health support.
\end{abstract}

\begin{IEEEkeywords}
social context, audio analysis, social anxiety
\end{IEEEkeywords}

\section{Introduction}

Social behavior and its variations across social contexts have been widely examined across multiple disciplines, from psychology to sociology \cite{liao2014understanding,phillips2018does} because of its centrality to understanding the human experience. The characteristics of social contexts, such as the number of interaction partners and perceived evaluative threats, are foundational to understanding communication and relationships, and provide insights into mental health, especially social anxiety disorder (SAD) \cite{asher2021m,toner2023wearable}. SAD, marked by an intense fear of social situations that prompt fears of negative evaluation from others, is highly prevalent and impairing but undertreated \cite{jefferies2020social}. Recognizing changes in social contexts likely to trigger social anxiety can guide detection of changes in mental status, pinpointing opportune moments for interventions \cite{nahum2018just}.

\begin{figure}
    \centering
    \includegraphics[width=0.9\columnwidth]{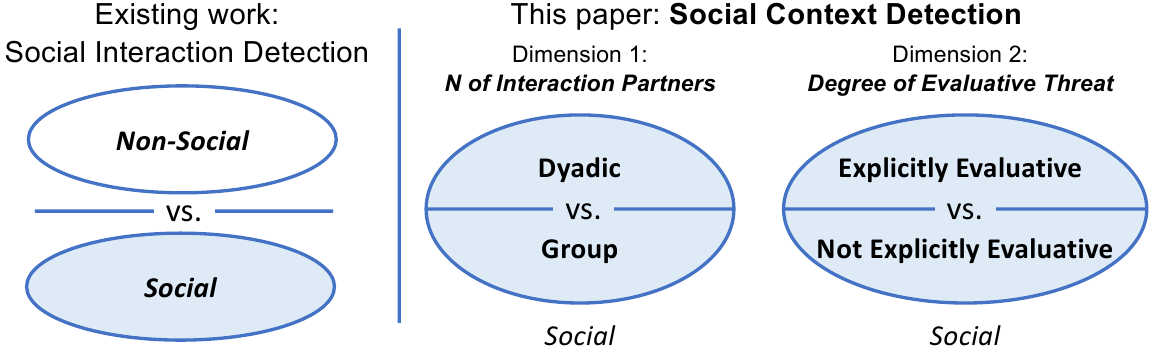}
    \caption{Comparison between existing work and our proposed social context recognition focusing on the number of interaction partners and degree of evaluative social threat.}
    \label{fig:intro}
\end{figure}

While self-reported data and laboratory-based studies have historically been instrumental in providing insights into social interactions and their contextual characteristics \cite{baumeister2007psychology,shiffman2008ecological,vazire2008knowing}, these methods often come with inherent limitations, being both burdensome and impractical for scalable and real-time applications \cite{olguin2008social,eagle2009inferring}. This limitation is particularly pertinent in the context of social anxiety, where the immediate and dynamic nature of interactions plays a crucial role. Advanced sensing technology, such as acoustic sensing available on smartphones and smartwatches, has emerged as a promising alternative, demonstrating potential in detecting social interactions \cite{katevas2019finding,liangdataset} (as illustrated on the left side of Figure \ref{fig:intro}) and assessing mental status \cite{salekin2018weakly,wang2022personalized}. This technology opens new avenues for analyzing the nuances of social interactions, like the number and type of interaction partners, which are critical in understanding the anxiety-provoking aspects of social contexts. However, the specific nuances within social interactions (e.g., the number of interaction partners) that determine their level of anxiety-provocation and the subtle changes in speech patterns associated with these contexts remain largely unexplored (see Figure 1 \ref{fig:intro}). This gap highlights the need for a focused approach that leverages the capabilities of advanced acoustic sensing to decipher these intricate aspects of social interactions to ultimately advance understanding of contextual influences on potential interventions for social anxiety.

In this paper, we leverage ambient audio segments to passively detect social contexts relevant to social anxiety, around two important dimensions: 1) interaction group size -- dyadic vs. group, and 2) the degree of evaluative threat -- explicitly evaluative vs. not-explicitly-evaluative (we assume that even when we are not giving direct evaluation instructions, there can be fears of evaluation, especially for socially anxious individuals; hence, we use the term not-explicitly-evaluative). Our study involved a series of Zoom-based virtual social interactions with N=52\footnote{The study initially involved 54 participants; however, the conversations of two individuals were not successfully recorded due to technical difficulties.} undergraduate students, among whom the majority (N=45) were high in socially anxiety symptoms (thereby ensuring our social threat context manipulations would be relevant). Data from four distinct interactions was collected: explicitly evaluative dyadic, not-explicitly-evaluative dyadic, explicitly evaluative group, and not-explicitly-evaluative group. We employ state-of-the-art audio2image techniques, which facilitate accurate detection of social context from spectrum and audio-feature embedded images using Convolutional Neural Networks (CNNs). Our methods, using audio segments within 7 seconds, achieved a 90\% accuracy in differentiating between dyadic and group interactions, and 83\% in identifying explicit evaluative threats, both measured under a 5-fold cross-validation (CV). More importantly, under a leave-one-group-out CV setting (iteratively leave out all data from a given set of participants to train and validate the models), the accuracy rates were 82\% and 77\% for the respective tasks.

While this study was primarily conducted with virtual interaction data from Zoom during the pandemic, its relevance extends beyond this context, reflecting the growing importance of virtual interactions in modern social life. The methodologies and insights we have developed are broadly applicable to everyday interactions, potentially captured through devices like smartphones and smartwatches. Our work underscores the potential of passive sensing and AI in understanding social contexts, opening avenues for context-aware, just-in-time mental health interventions. Initial tests with state-of-the-art models such as Transformers showed poor generalization on our limited dataset. Despite employing typical training techniques, these models did not perform well on the test sets, highlighting the limitations imposed by our sample size.

\section{Related Work}

While the potential of leveraging audio for detecting social interactions is evident, capturing the nuances of \textit{\textbf{social contexts}} requires a more detailed exploration. Research has demonstrated the feasibility of computer vision in identifying intense social interactions like real fights \cite{rota2015real}. Furthermore, studies have explored the physiological responses in evaluative contexts using wearable sensors \cite{toner2023wearable,wang2023detecting}. Efforts to understand social relations using audio-visual mediums have also come to the forefront, as seen in studies examining clinician-patient dynamics \cite{francesca2022lost}. In this study, we aim to capture the complexities of social contexts related to social anxiety, focusing solely on acoustic analysis.

Recent advancements have leveraged sensing signals to detect social interactions. Wang et al. \cite{wang2022personalized} utilized mobile sensing indicators to identify anxiety-relevant social contexts, including temporal phases and group sizes. The Electronically Activated Recorder (EAR) \cite{eagle2009inferring} has provided a means to capture extended audio data from conversations, while systems developed by researchers like Feng et al. \cite{feng2018tiles} optimize the detection of audio activity by calculating features during specific acoustic events. Acoustic sensing has emerged as a promising tool for understanding social interactions. Lane et al. \cite{lane2015deepear} employed deep learning methods to classify different acoustic events, and Liang et al. \cite{liang2023automated} demonstrated the potential of wearable technology in detecting face-to-face conversations.

Deep learning models such as Convolutional Neural Networks (CNNs) have shown considerable promise in audio processing \cite{hershey2017cnn}. Spectrograms, which convert audio signals into images, allow CNNs to discern patterns related to social contexts. The DeepInsight method \cite{sharma2019deepinsight} enhances CNN performance by mapping multidimensional audio features onto 2D planes, preserving relational structures within the data.

\section{Methods}

Leveraging a dataset collected from virtual social interactions conducted on Zoom, our analysis sought to detect key characteristics of social contexts that are often tied to social anxiety (i.e., number of interaction partners and degree of evaluative threat). This data collection included a range of sensors and features (e.g., physiological signals from wearable devices, self-reported survey responses, audio and video recordings from Zoom), but our goal for this paper was to learn whether audio on its own (selected because it is not burdensome, unlike self-report, and easily transportable, unlike video) could distinguish these important social context variations. The audio was processed through a combination of audio-to-image (audio2image) conversion and Convolutional Neural Networks (CNNs), leading to a model to detect specific social contexts. We used 26 acoustic features in our analysis, including Pitch, Jitter, Shimmer, Voice Breaks, Formant Frequencies, Formant Bandwidths, Intensity (dB), Harmonic Differences, HNR, MFCC, Spectral Centroid, Spectral Bandwidth, Spectral Contrast, Spectral Flatness, Spectral Rolloff, Zero Crossing Rate, RMS Energy, Chroma Features, Temporal Entropy, Autocorrelation, LPC Coefficients, Delta MFCC, Delta-Delta MFCC, Formant Amplitude, and Log Energy, which are widely recognized in acoustic analysis \cite{eyben2013recent}.

\subsection{Data Collection: Social Interactions Monitoring}

\subsubsection{Participants}

The study collected data (Zoom audio recordings) from a sample of N=52 undergraduate students. The majority of the participants (N=45) scored 34 or above on the Social Interaction Anxiety Scale (SIAS) \cite{mattick1998development}, a threshold indicating moderate to severe symptoms of social anxiety symptoms on a scale from 0 to 80, and N=7 subjects demonstrated low social anxiety (allowing for a full range of anxiety severity, but oversampling of individuals with heightened concerns about social threats). The study received approval from the Institutional Review Board (IRB) at a public university in the U.S., and all participants provided informed consent.

\subsubsection{Study Design and Procedure}

The study, which received full ethical approval and was under the supervision of a licensed clinical psychologist and researcher with expertise in anxiety disorders, involved the collection of audio data from virtual meeting recordings collected via Zoom. The data collection approach was purposefully crafted to manipulate various social contexts, including evaluative threat (explicitly evaluated or not explicitly evaluated), group size (dyadic vs. group conversation), and different anxiety phases (anticipatory, concurrent, and post-event). In this study, we focus on audio analysis across different evaluative threats and group sizes. 

Specifically, participants took part in four distinct social experiences\footnote{For some study sessions, there was no time to complete all interactions, causing different number of samples.}, i.e., dyadic and group conversations with explicit or not explicit evaluative threat, after a baseline phase where participants watched a neutral video alone. In detail:

\begin{itemize}
    \item \textbf{Dyadic Conversations} (twice with different levels of evaluative threat): Participants engaged in one-on-one conversations lasting 4 minutes each (e.g., a conversation between P01 and P02).
    \item \textbf{Group Conversations} (twice with different levels of evaluative threat): Conversations involved 3 to 6 participants and lasted 6 minutes each (e.g., a conversation including P01 to P06), with the number of participants contingent upon attendance.
\end{itemize}

Among the dyadic and group conversations, two levels of evaluative threat were manipulated as follows:

\begin{itemize}
\item  \textbf{Explicit Evaluative Threat}: In one dyadic and one group interaction, participants were explicitly informed beforehand that their performance would be evaluated by their conversational partners post-conversation.
\item \textbf{Non-Explicit Evaluative Threat}: For the other dyadic and group interactions, participants were informed that their performance would not be assessed by their conversational partners.
\end{itemize}

We provided each participant with random, unique conversation prompts, such as ``if you won a million dollars, how would you spend the money and why?'', for each interaction to initiate the conversation directions. Throughout the entire Zoom session, video recordings were made, with the start and end points of each social interaction noted to segment the audio data. It is important to note that, the order of the four social experiences was shuffled to ensure that the responses and interactions of participants were not influenced by the sequence of the experiences.

\begin{table}[t]
\centering
\small
\caption{Summary of the dataset collected.}
\begin{tabular}{lc}
\toprule
\textbf{Data} & \textit{Quantity} \\ 
\midrule
\textbf{Subjects} & \textbf{52}   \\ 
\quad Male &    12      \\
\quad Female &    40      \\
\textbf{Interactions Recordings} &   \textbf{65}      \\
\quad Evaluative Dyadic &    25 (4 min)      \\
\quad Not-Explicitly-Evaluative Dyadic &   23 (4 min)   \\
\quad Evaluative Group &     9 (6 min)  \\
\quad Not-Explicitly--Evaluative Group &    8  (6 min)    \\
\bottomrule
\end{tabular}
\end{table}

\begin{figure*}
    \centering
    \includegraphics[width=1\textwidth]{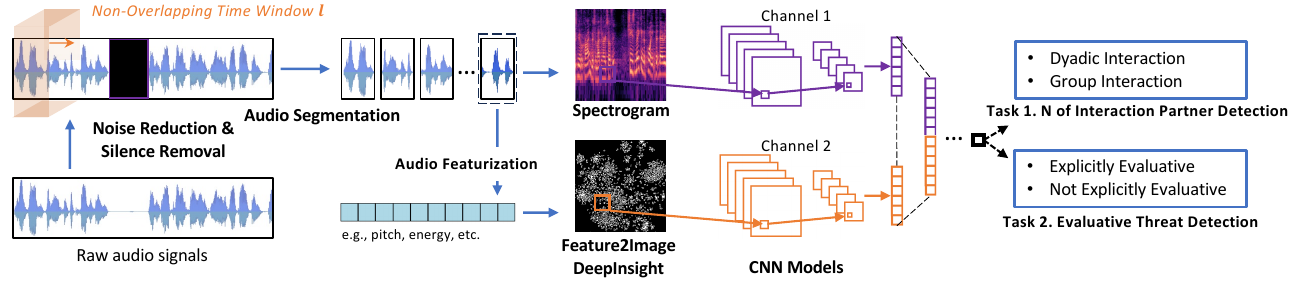}
    \caption{Illustration of the multi-channel CNN framework with audio-to-image processing for social context detection.}
    \label{fig:model-figure}
\end{figure*}

\subsection{Social Context Detection Framework}

Our framework, shown in Figure \ref{fig:model-figure}, detects social contexts from audio segments by converting pre-processed audio clips into multi-channel image representations. Of note, this approach avoids more `state-of-the-art' pre-trained Transformer-based methods like Audio spectrogram transformer \cite{gong2021ast} which typically have large model parameters, was to maintain a compact model scale and enhance flexibility for downstream deployment on edge devices with resource constraints.

\subsubsection{Audio Data Preprocessing}

We implemented noise-reduction techniques and eliminated segments with prolonged silence for clarity. Using Spectral Subtraction and short-term Fourier transform (STFT), we subtracted the estimated noise spectrum from the signal, focusing on low-energy frames. Silence removal was achieved by classifying frames with energy levels below 0.7 times the average as silence. Next, we segmented audio data into distinct, non-overlapping units of length $l$, treating each as an individual sample. Each sample was paired with its respective social context label.

% \subsubsection{Noise Reduction and Silence Removal}

% To ensure clarity, we used Spectral Subtraction for noise reduction and removed prolonged silence by computing frame energy. Frames with energy below 0.7 times the average were classified as silence and eliminated.

% \subsubsection{Audio Segmentation}

% We segmented audio data into distinct units of variable length $l$, treating each segment as an individual sample. Each sample was paired with its social context label, optimizing $l$ to capture the necessary contextual information for accurate classification.

\subsubsection{Audio-to-Image Transformation}

The audio segments are transformed into two unique image types - the Spectrogram and the DeepInsight \cite{sharma2019deepinsight} Feature2Image - to enable CNN-based social context classification. These transformations extract complex frequency and audio feature domain characteristics, thereby addressing a gap in traditional audio processing methodologies. The spectrogram transformation visualizes time-dependent frequency content, enabling the model to discern social context-related spectral patterns. The DeepInsight Feature2Image transformation overcomes spectrogram limitations by mapping multi-dimensional audio features to a 2D image space, preserving relational intricacies and facilitating higher-level abstractions for the CNN.

\textbf{Spectrogram:} Spectrograms translate one-dimensional time-series audio signals into two-dimensional images, capturing variations in frequency over distinct time intervals. Spectrograms preserve the time-series nature of the audio data, a factor critical to our task of social context detection. This form of representation, merging frequency and time information, adeptly captures voice variations instrumental for social context detection. The visualization ability of these time-frequency representations not only assists the CNN model but also provides valuable insights into the temporal context of these variations, thereby enriching our understanding of the underlying social context.

\textbf{DeepInsight Feature2Image:} The DeepInsight transformation technique leverages a multi-step process to convert audio feature data into a format suitable for a CNN \cite{sharma2019deepinsight}. It employs dimensionality reduction, using the t-SNE technique, to map each feature vector onto a 2D plane. This method retains the inherent structure and relationships within the original high-dimensional dataset. The resultant 2D map is a spatial configuration of points, each corresponding to a feature from the original data, with their positions indicative of their relational structure. The next phase involves defining the image frame. Using a convex hull algorithm, a rectangle is drawn that encompasses all the 2D points. Feature values that overlap are averaged, balancing information retention with hardware constraints and the number of features. Finally, 2D Cartesian coordinates of the features are translated and mapped into defined pixel positions, producing an image where the intensity of each pixel corresponds to the feature value, and the pixel's position represents the structural relationship of that feature to others. By enabling a CNN to capitalize on the structural and relational information among audio features encapsulated in an image format, the DeepInsight process enhances the classification performance for social context determination.

\subsection{Multi-Channel CNN-Based Classification}

Leveraging the strength of a multi-channel CNN to process concatenated grayscale images of the Spectrogram and the DeepInsight feature-embedding, we encompass perspectives on both time series frequency changes and audio feature variations.Despite being a single input image, our strategy treats each of the grayscale transformations as distinct channels, effectively preserving their individual characteristic information. The decision to use grayscale over color is backed by the assertion that the essential information in both the Spectrogram and the DeepInsight Feature2Image lies in pixel intensities rather than color variations. Our CNN architecture is designed to handle a dual-channel input. The first layer is an input layer configured to accept the two distinct channels. Following this, the network comprises three convolutional layers for each channel. Each convolutional layer is equipped with a ReLU activation function and is followed by a max pooling layer to reduce spatial dimensions and enhance feature extraction. Post these layers, a flattening step is implemented to convert the 2D feature maps into a 1D vector. This flattened vector is then passed through a dense layer and culminates in a softmax output layer, classifying the input into the relevant categories. To address hyperparameters such as convolution layers and filter sizes, we tuned these hyperparameters by applying Bayesian optimization techniques to ensure the optimal performance.

\section{Results}

In this section, we assess the proposed methods through two distinct evaluation configurations: 5-fold CV (sample-wide), where individual data may be part of both training and test sets, and leave-one-group-out CV (individual-specific), segregating the training and test sets into separate subject samples. These methods allow for a nuanced examination of the models' robustness and adaptability. Furthermore, we delve into the optimal audio sensing windows $l$ to gauge the efficacy of audio signals in detecting the two types of social contexts.

\subsection{Experiment Settings}

In our research, we investigate the effectiveness of our multi-channel CNN architecture against several baseline models. These baselines include two distinct CNN architectures, a Multi-layer Perceptron (MLP), a Random Forest, and a Decision Tree. 
The choice of these baseline models is grounded in their widespread acceptance and proven effectiveness in audio classification and similar signal-processing tasks. The CNN architectures are known for handling image-like data structures, making them suitable for our spectrogram and DeepInsight-transformed feature matrix images, respectively. The CNN architectures independently process the spectrogram and DeepInsight-transformed feature matrix images, each utilizing sequential 2D convolutional layers and max-pooling for spatial down-sampling. The MLP, Random Forest, and Decision Tree can handle traditional quantitative feature data. These models, widely used in various classification tasks, offer a robust baseline to evaluate the comparative advantage of our proposed CNN approach. 
The baseline models are trained on quantitative audio features of each audio segment. This diversity in baseline models allows for a comprehensive analysis of the strengths and limitations of different algorithmic approaches under similar conditions.

 In contrast, our multi-channel CNN architecture combines grayscale versions of the spectrogram and feature matrix images into a single input, processed in separate channels. By executing independent convolution and pooling stages for each channel before flattening and merging the derived features, this model captures the complementary information from both frequency and audio feature domains. The overarching objective is to ascertain whether this approach offers significant advantages over traditional audio classification methods.

To conduct holistic validation results, we trained models on samples segmented by the window length $l$. These models were evaluated using both a 5-fold cross-validation (CV) and a leave-one-group-out CV (LOGOCV), respectively. In the context of our study, the 5-fold CV spans the entire sample, meaning an individual's data can appear in both the training and test sets, showing a partially personalized model assessment because the model is trained and evaluated on different subsets of the same individuals' data, allowing it to adapt to individual variances. On the other hand, the LOGOCV is more challenging, where the training and test sets encompass distinct subject samples. To be more detailed, given our setup where participants were organized into Zoom meetings with the group including four to six members (e.g., P01-P06), the LOGOCV approach implies that we train the model using data from all groups except one and then validate using the left-out group. This simulates a real-world scenario where the trained model is applied to unseen, new individuals. The dual application of these CV methods provides multiple evaluations of the robustness and versatility of our models.

Throughout the experiments, we utilized evaluation metrics such as balanced accuracy and macro F1-score. Under class-imbalanced binary classifications, these metrics, by treating both target classes with equal weighted importance, retained a consistent baseline mean of 50\% regardless of the class imbalance.
\begin{equation} \label{balanced_accuracy}
\textit{Balanced Accuracy} = \frac{1}{2} \left( \textit{recall} + \textit{specificity} \right)
\end{equation}
\begin{align}
\textit{F1}_{\text{class\_i}} &= 2 \cdot \frac{\textit{precision}_{\text{class\_i}} \times \textit{recall}_{\text{class\_i}}}{\textit{precision}_{\text{class\_i}} + \textit{recall}_{\text{class\_i}}} \notag \\
\textit{Macro F1} &= \frac{1}{2} \cdot (\textit{F1}_{\text{class\_1}} + \textit{F1}_{\text{class\_2}})
\end{align}

\begin{table}[t]
\centering
\caption{Results of 5-fold-cross validation evaluation indicated by Balanced Accuracy and Macro-Weighted F1-Score. Models with ``+ Features'' utilize audio features as input, while models with ``+ DeepInsight'' and ``+ Spectrogram'' are trained with the two representations, respectively.}
\resizebox{\columnwidth}{!}{%
\begin{tabular}{lcccc}
\toprule
\multirow{2}{*}{\textbf{Methods}} & \multicolumn{2}{c}{\textbf{N of Partners}} & \multicolumn{2}{c}{\textbf{Evaluative Threat}} \\ 
\cmidrule(lr){2-3} \cmidrule(lr){4-5}
 & Accuracy & F1-score & Accuracy & F1-score \\ 
\midrule
Random Baseline & 0.50$\pm$0.00 & 0.50$\pm$0.00 & 0.50$\pm$0.00 & 0.50$\pm$0.00 \\ 
\midrule
Random Forest + Features & 0.83$\pm$0.01 & 0.84$\pm$0.02 & 0.80$\pm$0.01 & 0.79$\pm$0.01 \\
MLP + Features & 0.81$\pm$0.02 & 0.80$\pm$0.01 & 0.76$\pm$0.01 & 0.75$\pm$0.01 \\
Decision Tree + Features & 0.71$\pm$0.02 & 0.71$\pm$0.02 & 0.73$\pm$0.02 & 0.74$\pm$0.01 \\ 
\midrule
CNN + DeepInsight & 0.74$\pm$0.02 & 0.76$\pm$0.01 & 0.74$\pm$0.01 & 0.73$\pm$0.02 \\
CNN + Spectrogram & 0.87$\pm$0.02 & 0.86$\pm$0.03 & 0.85$\pm$0.02 & 0.85$\pm$0.02 \\
\midrule
\textit{(Ours)} Multi-Channel CNN & \textbf{0.90$\pm$0.02} & \textbf{0.86$\pm$0.03} & \textbf{0.83$\pm$0.02} & \textbf{0.87$\pm$0.01} \\ 
\bottomrule
\end{tabular}
}
\label{table:results_5_folds}
\end{table}

\begin{table}[t]
\centering
\caption{Results of Leave-One-Group-Out cross validation indicated by Balanced Accuracy and Macro-Weighted F1-Score.}
\resizebox{\columnwidth}{!}{%
\begin{tabular}{lcccc}
\toprule
\multirow{2}{*}{\textbf{Methods}} & \multicolumn{2}{c}{\textbf{N of Partners}} & \multicolumn{2}{c}{\textbf{Evaluative Threat}} \\ 
\cmidrule(lr){2-3} \cmidrule(lr){4-5}
 & Accuracy & F1-score & Accuracy & F1-score \\ 
\midrule
Majority Baseline & 0.50$\pm$0.00 & 0.50$\pm$0.00 & 0.50$\pm$0.00 & 0.50$\pm$0.00 \\ 
\midrule
Random Forest + Feature & 0.65$\pm$0.01 & 0.63$\pm$0.01 & 0.62$\pm$0.01 & 0.59$\pm$0.01 \\
MLP + Feature & 0.68$\pm$0.01 & 0.70$\pm$0.01 & 0.64$\pm$0.02 & 0.64$\pm$0.02 \\
Decision Tree + Feature & 0.59$\pm$0.01 & 0.60$\pm$0.01 & 0.54$\pm$0.02 & 0.52$\pm$0.02 \\ 
\midrule
CNN + DeepInsight & 0.66$\pm$0.01 & 0.64$\pm$0.01 & 0.62$\pm$0.01 & 0.73$\pm$0.01 \\
CNN + Spectrogram & 0.71$\pm$0.02 & 0.72$\pm$0.03 & 0.72$\pm$0.02 & 0.70$\pm$0.02 \\
\midrule
\textit{(Ours)} Mutli-Channel CNN & \textbf{0.82$\pm$0.02} & \textbf{0.82$\pm$0.02} & \textbf{0.77$\pm$0.01} & \textbf{0.78$\pm$0.02} \\ 
\bottomrule
\end{tabular}}
\label{table:results_lopo}
\end{table}

\subsection{Experiment Results}

\subsubsection{Model Efficiency}

This section presents the performance results obtained from the various models under two distinct evaluation setups: 5-fold CV and LOGO CV.

\textbf{5-Fold Cross Validation}

Table \ref{table:results_5_folds} showcases the results of models evaluated under the 5-fold CV, assessing the models based on Balanced Accuracy and Macro-Averaged F1-score for two primary tasks: Number of Interaction Partners and Evaluative Threat detection.

Given the evaluation metrics used, treating both target classes with equal weighted importance for each fold of the validation establishes a baseline performance (labeled as ``Majority Mean'' in Table \ref{table:results_5_folds}) where always guessing the majority class results in scores of 50\% for both metrics. Traditional machine learning models such as Random Forest, MLP, and Decision Tree, which use audio features, yielded strong results, indicating the potential utility of using audio features for the task. Among the CNN variants, the multi-channel approach, which was expected to perform more effectively than traditional approaches (see Experiment Settings), stood out with an accuracy score of 90.0\% and an F1-score of 86\% for Number of Interaction Partners detection, and 83\% accuracy and 87\% F1-score for Evaluative Threat detection, respectively, clearly outperforming the other approaches tested. This suggests that leveraging both spectrogram and DeepInsight representations concurrently within a CNN offers an advantage over the individual use of these features or traditional feature-based models.

\textbf{Leave-One-Group-Out}

Table \ref{table:results_5_folds} presents the evaluation results using the LOGOCV setup. In this stringent test, the training and test sets comprise completely separate participants, offering broader generalizability of the model results.

Similar to the 5-fold cross-validation results, traditional machine learning models fared reasonably well but were outperformed by CNN-based methods (83\% accuracy and 82\% F1-score for Number of Interaction Partners and 77\% accuracy and 78\% F1-score for Evaluative Threat detection, respectively). This underscores the robustness and versatility of our model, particularly when applied to unseen subjects.

\begin{figure}[t]
\centering
\subfigure[Sample-wide 5-fold CV]{%
  \includegraphics[width=0.995\columnwidth]{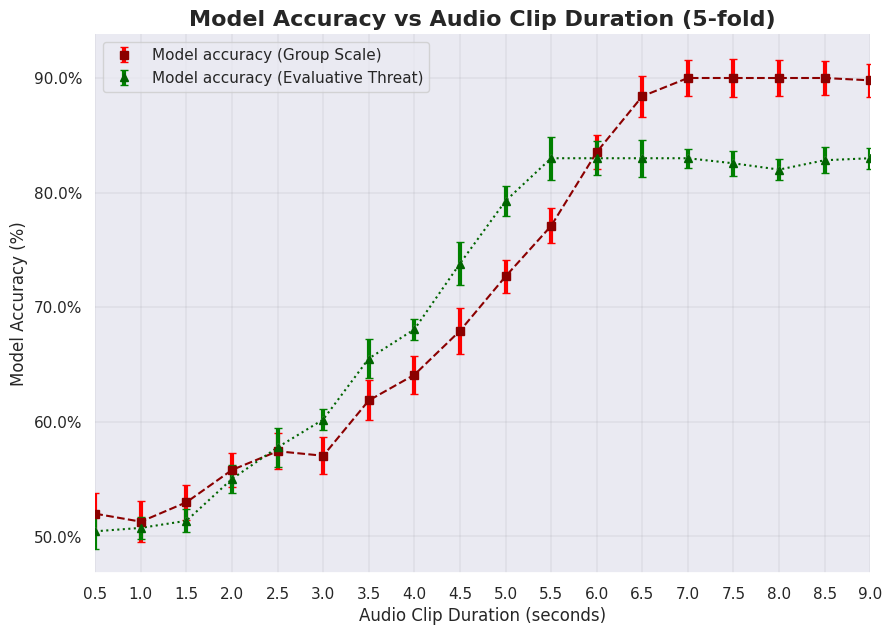}%
  \label{fig:window-5-fold}%
}
\subfigure[Leave-one-group-out CV]{%
  \includegraphics[width=0.995\columnwidth]{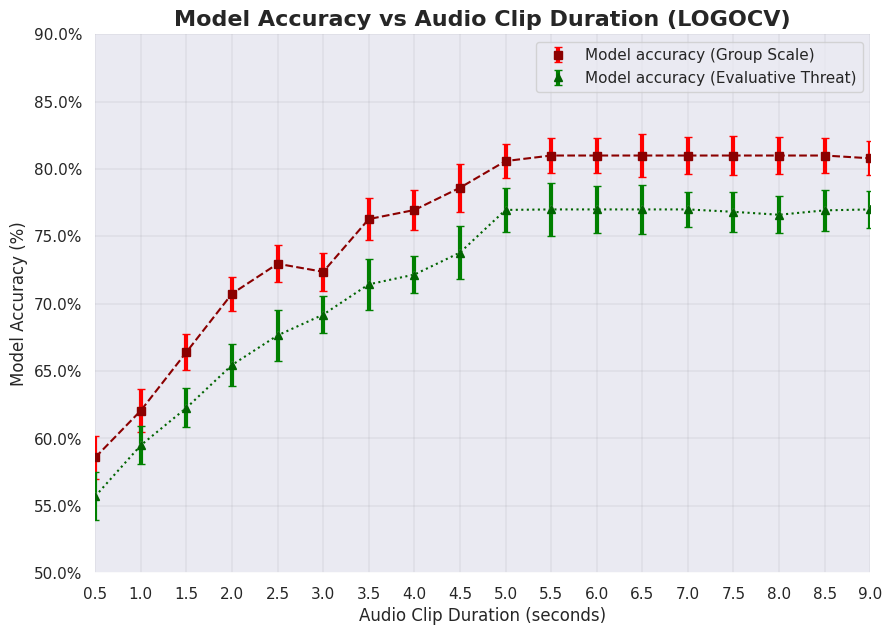}%
  \label{fig:window-logocv}%
}
\caption{Performance of the multi-channel CNN models trained by varying length $l$ of audio segments under 5-Fold CV (a) and LOGOCV (b) paradigms. (The error bars indicate the standard deviations of the performance of the models trained for 10 times per point).}
\label{fig:optimal-window}
\end{figure}

\subsubsection{Optimal Sensing Window}

To understand the nuanced relationship between the detection performance of social contexts and the segment length $l$ of audio, we trained models on audio segments with lengths ranging from 0.5 to 9 seconds. The balanced accuracy served as the exclusive metric.

As shown in Figure \ref{fig:window-5-fold}, for sample-wide 5-fold CV: When distinguishing between dyadic and group contexts, the model's accuracy started near the baseline of 50\% with 0.5-second audio. It showed a sharp increase, reaching its zenith at 90\% for an audio length of 7.0 seconds. For identifying evaluative threats, the model's accuracy peaked at 83\% with a segment length of 5.5 seconds. See Figure \ref{fig:window-logocv}, regarding the LOGO CV: The differentiation of number of interaction partners achieved its highest accuracy of 82\% at a segment length of 5.5 seconds. The task of detecting evaluative threats reached its optimal performance, with an accuracy of 77\% for segments lasting 5.0 seconds.

These results highlight the significance of fine-tuning the audio segment length for optimal detection performance and also illustrate the dynamic trade-offs between audio length and model performance of social context detection.

\section{Discussion}

\subsection{Technical Implications}

% Feasibility and robustness of proposed methods
Our study demonstrates strong model performance, particularly after integrating information from multiple channels, including time series frequency changes and audio feature variations. This multidimensional audio2image approach was effective at capturing nuanced aspects of social interactions.

% how are the results aligned with and promoting existing work.
Prior work has found that using physiological cues like heart rate variability and motion to detect social contexts yielded accuracies of only 50\% (suggesting limited effectiveness) in social evaluation detection and 61\% in distinguishing dyadic from group interactions \cite{toner2023wearable}. Another line of research leveraged acoustic sensing to generally detect whether or not a face-to-face social interaction was occurring, achieving 80.4\% accuracy \cite{liangdataset}. The present work showcases the advantages of acoustic sensing to more accurately detect more nuanced social contexts.

% mention the 5-fold and lopocv configuration again
The discrepancies observed between the outcomes of the 5-fold CV and the LOGOCV are informative. In particular, the LOGOCV poses a more challenging test, probing the model's generalizability across unique participant subsamples. 

% further technical implications and promising points for future 
The efficacy of our audio-based methods raises intriguing questions about its applicability to other social interaction contexts (e.g., conversations between therapists and clients; workspace interactions). It will be important to directly test generalizability to other virtual and in-person interactions. Additionally, future endeavors might contemplate supplementing our acoustic data with other sensed inputs, like visual or physiological markers. At the same time, the model's success based on only short audio segments suggests an efficient avenue for real-time detection and processing, which can be readily integrated into everyday edge devices like smartwatches and smartphones, and other digital communication platforms (e.g., Telehealth) to bolster their context-awareness. The minimal audio data needed, combined with the possibility of customization to account for individual speech nuances and cultural variations, suggests a direction towards a personalized user experience with context-awareness fine-tuning.

\subsection{Social Impact and Applications}

% enhancing social interaction understanding
With the ability to more precisely characterize social contexts from audio segments, our tool has potentially important implications for fields like clinical and social psychology. 
This advancement not only aligns with, but also extends established intervention approaches, like cognitive behavior therapy.
By understanding the varying dynamics and contexts of social interactions, therapists and digital health apps could potentially tailor their interventions. For example, instead of relying on client's retrospective self-report to understand how a social interaction went for a client, a therapist could directly learn how turn-taking (i.e., balancing speaking and listening) played out in a real-world dyadic interaction, and recommend different approaches to enhance interpersonal effectiveness accordingly. In contrast, during group conversations, advice might center on tactics for entering a discussion when multiple people are actively participating, or ways to navigate more vs. less evaluative interaction contexts. 

% Broadening the Application Spectrum
Beyond implications for social anxiety and threat detection, distinguishing among small versus larger group interactions has many applications for studying group dynamics and formation, social networks and hierarchies, etc. More generally, this approach contributes to a more nuanced understanding of social interactions, raising new possible applications in the context of passive sensing and AI's role in Just-in-Time Adaptive Interventions (JITAIs) \cite{nahum2018just}.

% ethical and privacy
While our research suggests promising future applications (assuming these results replicate and extend to other samples and social contexts), ethical and privacy concerns remain paramount given the sensitive and identifying nature of audio data. The efficacy of detecting social contexts from transformed images of short audio segments (i.e., about 1-5 seconds) underscores a promising trade-off between utility and privacy. This short duration not only minimizes the amount of verbal content captured, thereby reducing privacy risks and concerns related to verbal content leakage, but also allows for computationally low-cost deployment of the model on edge devices, which ensures that raw audio data are neither stored nor transmitted beyond the device. By doing so, the privacy risks can be greatly minimized, but a critical next step will be more qualitative work with participants to better understand their views toward monitoring of audio and the real-world conditions where ongoing monitoring would feel acceptable and non-intrusive.

\subsection{Limitations and Future Work}

While our study presents key advances, there are of course also limitations which point to opportunities for future work:

\textbf{Study Constraints and Settings:} Our study was conducted during the pandemic, which influenced our data collection strategies and methods. Specifically, data was collected through Zoom due to social distancing restrictions. This digital environment, while highly relevant to current conditions, may not capture the full spectrum and nuances of real-world social interactions. Also, our measured social interactions were limited to experimentally designed, binary social contexts (dyadic vs. group and evaluative vs. not-explicitly-evaluative) on Zoom, but everyday interactions can vary on many more, complex, dynamic variables (e.g., whether the individual is familiar with the interaction partners). Moving forward, it will be valuable to assess in-person social interactions across a wide variety of contexts and interaction partners (e.g., using microphones in mobile devices, assuming appropriate permissions have been obtained).

\textbf{Participant Diversity:} The sample predominantly consisted of English-speaking, cisgender female undergraduate students aged 18-22 years. Future studies should include samples with a broader age range, cultural and language backgrounds, and other demographic factors.

\textbf{Implementation and Integration:} The potential extensions and implications of our proposed method are considerable, particularly for integration into next-generation sensing systems such as mobile-based digital health systems \cite{wang2022personalized}. However, several questions about its practical application, real-time processing capabilities, and scalability across diverse settings persist. Implementing an audio-based assistant system brings forward privacy and ethical considerations that are critical to address. To address this, future research could explore incorporating privacy-preserving techniques and ethical protocols into social context detection systems. For instance, the proposed methods that necessitate short audio segments with minimal speech content for computations could leverage edge devices for processing, without needing to gather data. The approach, which involves extracting audio features at the edge device, deleting the audio files thereafter, and completing the detection task based on features, may also offer a viable solution while addressing privacy concerns.

\section{Conclusion}

In this paper, we proposed to detect nuanced social contexts using ambient audio segments, highlighting the efficacy of our audio2image techniques in conjunction with CNNs. The obtained accuracies, under both the 5-fold and leave-one-group-out CV scenarios, show promise for a meaningful advance in social interaction monitoring. While our data source, being Zoom-based due to the pandemic, presents some constraints, we anticipate (but need to test) that the methodology used will extend to a broad range of real-world settings. The promise of passive sensing combined with AI models, as showcased in our results, emphasizes the potential for accurate, real-time social context detection. 

\section*{Ethical Impact Statement}

All study procedures were approved by the Institutional Review Board (UVA-SBS IRB \#3004) of a U.S. university and conducted under the supervision of a licensed clinical psychologist and researcher with expertise in anxiety disorders. Participants provided informed consent and were made fully aware of the study's purpose, the nature of the data collection, and their rights within the research. Moreover, at the conclusion of the study session, participants were asked to confirm they consented for us to analyze their audio and video data. 

\textbf{Data Privacy}: In this study, Zoom recording data was collected using password-protected computers by research assistants and securely stored on a high-security data server only accessible via high security virtual private network (VPN). Importantly, the approach taken in the present study minimizes the risk of data misuse while still enabling the identification of social contexts relevant to social anxiety by: 1) training detection models that are easily deployable on ubiquitous devices (like smartphones and smartwatches) to ensure that raw audio data are not stored or transmitted beyond the device, and 2) only requiring seconds-long audio.  We advocate for transparency in any subsequent research or applications that keep the users fully aware of how data will be collected and used, and allow users to opt in or out freely.

\textbf{Generalizability}: From a technical perspective, this study employed leave-one-group-out cross validation to evaluate the trained model's generalizability to unseen individuals. However, we also acknowledge that the present findings, while promising, are based on data collected from college students in controlled, virtual interactions, and their application to more diverse samples and real-world settings requires further validation. We regard this study as a preliminary, proof-of-concept step in an important social context (given college students have high rates of social anxiety, and virtual environments have become an integral aspect of daily life for many in the post-COVID world). It will also be important to conduct research with people who hold different identities and are engaged in different contexts to determine generalizability of these results. 

Overall, this work contributes to the broader field of affective computing research by offering new insights into how social contexts can be detected and understood through non-intrusive ambient audio sensing. We envision that this work can pave the way for practical tools (e.g., context-aware JITAI systems that can sense when and where individuals require support most and provide personalized interventions tailored to their specific contexts) that improve the lives of those with social anxiety, while adhering to ethical and data privacy standards.

\bibliographystyle{IEEEtran}
\bibliography{reference}

% Generated by IEEEtran.bst, version: 1.14 (2015/08/26)
\begin{thebibliography}{10}
\providecommand{\url}[1]{#1}
\csname url@samestyle\endcsname
\providecommand{\newblock}{\relax}
\providecommand{\bibinfo}[2]{#2}
\providecommand{\BIBentrySTDinterwordspacing}{\spaceskip=0pt\relax}
\providecommand{\BIBentryALTinterwordstretchfactor}{4}
\providecommand{\BIBentryALTinterwordspacing}{\spaceskip=\fontdimen2\font plus
\BIBentryALTinterwordstretchfactor\fontdimen3\font minus
  \fontdimen4\font\relax}
\providecommand{\BIBforeignlanguage}[2]{{%
\expandafter\ifx\csname l@#1\endcsname\relax
\typeout{** WARNING: IEEEtran.bst: No hyphenation pattern has been}%
\typeout{** loaded for the language `#1'. Using the pattern for}%
\typeout{** the default language instead.}%
\else
\language=\csname l@#1\endcsname
\fi
#2}}
\providecommand{\BIBdecl}{\relax}
\BIBdecl

\bibitem{liao2014understanding}
Y.~Liao, S.~Intille, J.~Wolch, M.~A. Pentz, and G.~F. Dunton, ``Understanding
  the physical and social contexts of children’s nonschool sedentary
  behavior: an ecological momentary assessment study,'' \emph{Journal of
  Physical Activity and Health}, vol.~11, no.~3, pp. 588--595, 2014.

\bibitem{phillips2018does}
K.~T. Phillips, M.~M. Phillips, T.~L. Lalonde, and M.~A. Prince, ``Does social
  context matter? an ecological momentary assessment study of marijuana use
  among college students,'' \emph{Addictive Behaviors}, vol.~83, pp. 154--159,
  2018.

\bibitem{asher2021m}
M.~Asher, S.~G. Hofmann, and I.~M. Aderka, ``I’m not feeling it: Momentary
  experiential avoidance and social anxiety among individuals with social
  anxiety disorder,'' \emph{Behavior therapy}, vol.~52, no.~1, pp. 183--194,
  2021.

\bibitem{toner2023wearable}
E.~R. Toner, M.~Rucker, Z.~Wang, M.~A. Larrazabal, L.~Cai, D.~Datta,
  E.~Thompson, H.~Lone, M.~Boukhechba, B.~A. Teachman \emph{et~al.}, ``Wearable
  sensor-based multimodal physiological responses of socially anxious
  individuals across social contexts,'' \emph{arXiv preprint arXiv:2304.01293},
  2023.

\bibitem{jefferies2020social}
P.~Jefferies and M.~Ungar, ``Social anxiety in young people: A prevalence study
  in seven countries,'' \emph{PloS one}, vol.~15, no.~9, p. e0239133, 2020.

\bibitem{nahum2018just}
I.~Nahum-Shani, S.~N. Smith, B.~J. Spring, L.~M. Collins, K.~Witkiewitz,
  A.~Tewari, and S.~A. Murphy, ``Just-in-time adaptive interventions (jitais)
  in mobile health: key components and design principles for ongoing health
  behavior support,'' \emph{Annals of Behavioral Medicine}, vol.~52, no.~6, pp.
  446--462, 2018.

\bibitem{baumeister2007psychology}
R.~F. Baumeister, K.~D. Vohs, and D.~C. Funder, ``Psychology as the science of
  self-reports and finger movements: Whatever happened to actual behavior?''
  \emph{Perspectives on psychological science}, vol.~2, no.~4, pp. 396--403,
  2007.

\bibitem{shiffman2008ecological}
S.~Shiffman, A.~A. Stone, and M.~R. Hufford, ``Ecological momentary
  assessment,'' \emph{Annu. Rev. Clin. Psychol.}, vol.~4, pp. 1--32, 2008.

\bibitem{vazire2008knowing}
S.~Vazire and M.~R. Mehl, ``Knowing me, knowing you: the accuracy and unique
  predictive validity of self-ratings and other-ratings of daily behavior.''
  \emph{Journal of personality and social psychology}, vol.~95, no.~5, p. 1202,
  2008.

\bibitem{olguin2008social}
D.~O. Olgu{\'\i}n and A.~S. Pentland, ``Social sensors for automatic data
  collection,'' \emph{AMCIS 2008 Proceedings}, p. 171, 2008.

\bibitem{eagle2009inferring}
N.~Eagle, A.~Pentland, and D.~Lazer, ``Inferring friendship network structure
  by using mobile phone data,'' \emph{Proceedings of the national academy of
  sciences}, vol. 106, no.~36, pp. 15\,274--15\,278, 2009.

\bibitem{katevas2019finding}
\BIBentryALTinterwordspacing
K.~Katevas, K.~H\"{a}nsel, R.~Clegg, I.~Leontiadis, H.~Haddadi, and
  L.~Tokarchuk, ``Finding dory in the crowd: Detecting social interactions
  using multi-modal mobile sensing,'' in \emph{Proceedings of the 1st Workshop
  on Machine Learning on Edge in Sensor Systems}, ser. SenSys-ML 2019.\hskip
  1em plus 0.5em minus 0.4em\relax New York, NY, USA: Association for Computing
  Machinery, 2019, p. 37–42. [Online]. Available:
  \url{https://doi.org/10.1145/3362743.3362959}
\BIBentrySTDinterwordspacing

\bibitem{liangdataset}
D.~Liang, Z.~Xu, Y.~Chen, R.~Adaimi, D.~Harwath, and E.~Thomaz, ``A dataset for
  foreground speech analysis with smartwatches in everyday home environments,''
  in \emph{2023 IEEE International Conference on Acoustics, Speech, and Signal
  Processing Workshops (ICASSPW)}, 2023, pp. 1--5.

\bibitem{salekin2018weakly}
A.~Salekin, J.~W. Eberle, J.~J. Glenn, B.~A. Teachman, and J.~A. Stankovic, ``A
  weakly supervised learning framework for detecting social anxiety and
  depression,'' \emph{Proceedings of the ACM on interactive, mobile, wearable
  and ubiquitous technologies}, vol.~2, no.~2, pp. 1--26, 2018.

\bibitem{wang2022personalized}
Z.~Wang, H.~Xiong, J.~Zhang, S.~Yang, M.~Boukhechba, D.~Zhang, L.~E. Barnes,
  and D.~Dou, ``From personalized medicine to population health: a survey of
  mhealth sensing techniques,'' \emph{IEEE Internet of Things Journal}, vol.~9,
  no.~17, pp. 15\,413--15\,434, 2022.

\bibitem{rota2015real}
P.~Rota, N.~Conci, N.~Sebe, and J.~M. Rehg, ``Real-life violent social
  interaction detection,'' in \emph{2015 IEEE international conference on image
  processing (ICIP)}.\hskip 1em plus 0.5em minus 0.4em\relax IEEE, 2015, pp.
  3456--3460.

\bibitem{wang2023detecting}
\BIBentryALTinterwordspacing
Z.~Wang, M.~A. Larrazabal, M.~Rucker, E.~R. Toner, K.~E. Daniel, S.~Kumar,
  M.~Boukhechba, B.~A. Teachman, and L.~E. Barnes, ``Detecting social contexts
  from mobile sensing indicators in virtual interactions with socially anxious
  individuals,'' \emph{Proc. ACM Interact. Mob. Wearable Ubiquitous Technol.},
  vol.~7, no.~3, sep 2023. [Online]. Available:
  \url{https://doi.org/10.1145/3610916}
\BIBentrySTDinterwordspacing

\bibitem{francesca2022lost}
P.~Francesca, S.~M. Weldon, and A.~Lomi, ``Lost in translation: Collecting and
  coding data on social relations from audio-visual recordings,'' \emph{Social
  Networks}, vol.~69, pp. 102--112, 2022.

\bibitem{feng2018tiles}
T.~Feng, A.~Nadarajan, C.~Vaz, B.~Booth, and S.~Narayanan, ``Tiles audio
  recorder: an unobtrusive wearable solution to track audio activity,'' in
  \emph{Proceedings of the 4th ACM Workshop on Wearable Systems and
  Applications}, 2018, pp. 33--38.

\bibitem{lane2015deepear}
N.~D. Lane, P.~Georgiev, and L.~Qendro, ``Deepear: robust smartphone audio
  sensing in unconstrained acoustic environments using deep learning,'' in
  \emph{Proceedings of the 2015 ACM international joint conference on pervasive
  and ubiquitous computing}, 2015, pp. 283--294.

\bibitem{liang2023automated}
\BIBentryALTinterwordspacing
D.~Liang, A.~Zhang, and E.~Thomaz, ``Automated face-to-face conversation
  detection on a commodity smartwatch with acoustic sensing,'' \emph{Proc. ACM
  Interact. Mob. Wearable Ubiquitous Technol.}, vol.~7, no.~3, sep 2023.
  [Online]. Available: \url{https://doi.org/10.1145/3610882}
\BIBentrySTDinterwordspacing

\bibitem{hershey2017cnn}
S.~Hershey, S.~Chaudhuri, D.~P. Ellis, J.~F. Gemmeke, A.~Jansen, R.~C. Moore,
  M.~Plakal, D.~Platt, R.~A. Saurous, B.~Seybold \emph{et~al.}, ``Cnn
  architectures for large-scale audio classification,'' in \emph{2017 ieee
  international conference on acoustics, speech and signal processing
  (icassp)}.\hskip 1em plus 0.5em minus 0.4em\relax IEEE, 2017, pp. 131--135.

\bibitem{sharma2019deepinsight}
A.~Sharma, E.~Vans, D.~Shigemizu, K.~A. Boroevich, and T.~Tsunoda,
  ``Deepinsight: A methodology to transform a non-image data to an image for
  convolution neural network architecture,'' \emph{Scientific reports}, vol.~9,
  no.~1, p. 11399, 2019.

\bibitem{eyben2013recent}
F.~Eyben, F.~Weninger, F.~Gross, and B.~Schuller, ``Recent developments in
  opensmile, the munich open-source multimedia feature extractor,'' in
  \emph{Proceedings of the 21st ACM international conference on Multimedia},
  2013, pp. 835--838.

\bibitem{mattick1998development}
R.~P. Mattick and J.~C. Clarke, ``Development and validation of measures of
  social phobia scrutiny fear and social interaction anxiety,'' \emph{Behaviour
  research and therapy}, vol.~36, no.~4, pp. 455--470, 1998.

\bibitem{gong2021ast}
Y.~Gong, Y.-A. Chung, and J.~Glass, ``Ast: Audio spectrogram transformer,''
  \emph{arXiv preprint arXiv:2104.01778}, 2021.

\end{thebibliography}
\end{document}